\documentstyle[12pt,axodraw]{article}

\def\tid{\hspace{.3mm}}
\def\l{\lambda}

\def\D{{\cal D}}
\def\M{{\cal M}}

\def\r{\rho}
\def\s{\sigma}

\def\Sl#1{#1\!\!\!\!/\,}
\def\sl#1{#1\!\!\!/\,}
\def\wh#1{\widehat{#1}}
\def\wt#1{\widetilde{#1}}

\def\ol#1{\overline{#1}} 

\def\ts{\textstyle}
\def\d{\partial}
\def\m{\mu}
\def\n{\nu}

\def\e{\epsilon}

\def\F{{\cal F}}
\def\R{{\cal R}}

\def\be{\begin{equation}}
\def\ee{\end{equation}}
\def\beq{\begin{equation}}
\def\eeq{\end{equation}}
\def\bea{\begin{eqnarray}}
\def\eea{\end{eqnarray}} 
\def\beqa{\begin{equation}\begin{array}{l}}
\def\eeqa{\end{array}\end{equation}}

\def\eqn#1{(\ref{#1})}

\def\eqref#1{eq.~(\ref{eq:#1})}

  \def\g{\gamma}
 
\def\veps{\varepsilon}

\def\w{\omega}

\def\nn{\nonumber}

\def\mathscr{\mathcal}

\def\psib{\overline{\psi}}

\begin{document}

\thispagestyle{empty}
\begin{flushright}
\framebox{\small BRX-TH 469}\\
\end{flushright}

\vspace{.8cm}
\setcounter{footnote}{0}
\begin{center}
{\Large{\bf 
Massive Spin 3/2 Electrodynamics
    }
    }\\[10mm]

{\sc S. Deser~$\!^\sharp$, 
V. Pascalutsa~$\!^\flat$
and A. Waldron~$\!^\sharp$~\\[6mm]}

{\em\small ${}^\sharp$ 
Physics Department, Brandeis University, Waltham,
MA 02454, 
USA\\ {\tt deser,wally@brandeis.edu}}\\[5mm]

{\em\small ${}^\flat$ Department of Physics, Flinders University,
Bedford Park, SA 5042, Australia\\
{\tt phvvp@flinders.edu.au}}\\[5mm]

{\small (\today)}\\[1cm]

{\sc Abstract}\\
\end{center}

{\small
\begin{quote}

We study the  general non-minimally coupled charged massive spin~3/2
model both for its low energy phenomenological properties and for
its unitarity, causality and degrees of freedom behaviour.
When the model is viewed as an effective theory, 
its parameters (after ensuring
the correct excitation count) are related to physical characteristics, such
as the magnetic moment
$g$ factor, by means of low energy theorems.
We also provide the corresponding higher spin
generalisation. Separately, we consider 
both low and high energy unitarity, as well as 
the causality aspects of our
models. None  (including
truncated $N=2$ supergravity) is free of the
minimal model's acausality.

\bigskip

\bigskip


\end{quote}
}

\newpage





\section{Introduction}

Gauge interactions of 
massive (let alone massless)
relativistic higher spin fields 
constitute an ancient and difficult subject.
Whatever the formal problems these models encounter, {\it effective} 
higher spin theories must be constructible since
approximately localised higher spin particles exist. Such models should
achieve low energy consistency, and share some of the 
physical properties described by
their lower spin hadronic physics counterparts.

The proper framework for describing relativistic interactions is 
the action principle. In our present study of 
charged massive higher spins 
we seek only effective (rather than
renormalisable) actions, which can in general possess
dimensionful, non-minimal, couplings
beyond the
minimal prescription,
unique in first order systems,
$\partial_\mu\rightarrow\partial_\mu+ieA_\mu$. 
In particular, the associated coupling constants will determine
$2s+1$ intrinsic multipole moments of a spin $s$
particle (charge, magnetic dipole, quadrupole and
octupole for $s=3/2$).

The more formal properties, such as
unitarity and causality, of higher spin models
will in general also depend upon details of the 
non-minimal couplings. Some of the important affected issues
include: (i) A gyromagnetic ratio $g=2$ is required by
the optical and low energy theorems, at least for for pure electromagnetic
interactions~\cite{Weinberg:1970bu}, on the other hand, minimal
coupling implies $g=2/3$~\cite{Belinfante:1953}.   (ii) Tree
unitarity~\cite{Cornwall:1974km} requires
the non-minimal couplings of (truncated) $N=2$
supergravity~\cite{Ferrara:1992yc}.
(iii) Quantisation of the minimal
theory is problematic since the
fundamental canonical commutator becomes indefinite beyond a critical
value of the magnetic field~\cite{Johnson:1961vt}, or equivalently
the model exhibits acausal
propagation~\cite{Velo:1969bt}. However, if the 
minimal electromagnetic interactions are
extended to include   
gravity as 
obtained by dropping only the cosmological constant
term of $N=2$ anti-de Sitter supergravity~\cite{Freedman:1977aw},
causality is restored~\cite{Deser:1977uq}.
Unfortunately, this formally consistent model is unsuitable for
phenomenological applications, since it tunes
spin~3/2 and  Planck masses.

In this paper, we study the low energy behaviour, unitarity 
and causal consistency of general flat space 
non-minimal, non-derivative, 
couplings, concentrating on the massive charged spin~3/2 system,
the simplest theory subject to the array of higher spin
subtleties.
Of the five independent non-minimal terms linear in the field
strength,  two are eliminated by a simple degrees of 
freedom (DOF) consistency requirement. 
One of the three remaining
couplings does not contribute to 
photon
emission and Compton scattering 
but is included in our causality
analysis.
We employ low energy theorems (LETs) 
to identify the leading low energy non-minimal coupling as a 
magnetic moment interaction and compute the gyromagnetic ratio $g$
in terms of the parameters in the action.
The generalisation of this result to higher multipoles will be treated
elsewhere~\cite{Deser:2000}; its higher spin analogue
is given in Appendix~\ref{appA}.

The gyromagnetic ratio $g$ in our models is arbitrary and thus  
they are perfectly
suited for phenomenological applications where neither the pure
electromagnetic $g=2$ unitarity requirement nor tree unitarity
need apply. Loss of the latter 
merely signals the scale at
which the effective description ceases to be valid.

Study of causality yields a
negative result; like the minimal model, ours 
all permit acausal propagation for critical
electromagnetic fields. [In~\cite{Deser:1977uq} 
causality is preserved by taking gravity and electromagnetism
not merely external, but dynamical; 
we hope to include curved
space elsewhere~\cite{Deser:2000}.] Nevertheless, we will argue that for
perturbative processes, formulated in terms of free asymptotic
fields, neither high energy unitarity nor causality problems spoil the
validity of the models as a 
phenomenological tool.

In section~\ref{themodel} we
present the non-minimal models under
consideration and obtain the constraints required by
a correct DOF count. Section~\ref{low}
contains the verification of the LETs for the soft
photon vertex and Compton scattering, along with our gyromagnetic ratio
computation. A study of causality is
presented in section~\ref{causal}. In 
section~\ref{conclusion} we summarise and discuss our work.
The generalisation of our soft photon vertex
results and identification of the magnetic moment for higher spins is 
given in Appendix~\ref{appA} and the extension of our causality
analysis
to the most general non-minimal couplings is given in Appendix~\ref{appB}. 

\section{The Model}
\label{themodel}

We begin with the Lagrangian for the complex vector-spinor
Rarita--Schwinger
field $\psi_\mu$,
\be
{\cal L}\,=\,-\,\psib\,^\mu \g_{\m\n\r}\,\D^\n\psi^\r-
\frac{ie}{m}\,\psib{}_\m\,
\F^{\m\n}\psi_\n\, ;
\label{action}
\ee
\bea
\psib{}_\m\,\F^{\m\n}\psi_\n&\equiv\!\!\!&\phantom{+\,}
l_1\,\psib{}_\mu F^{\mu \nu}\psi_\nu
\,+\,
l_2\,\psib{}_\mu \wh F \,\psi^\mu\nn\\&&
+\,
l_3\,F^{\m \n}\,[\psib{}_\mu  \g_\n \,\g.\psi+\psib.\g \,\g_\m \psi_\n]
\,+\,
l_4\,\psib.\g\, \wh F\, \g.\psi\nn\\&&
+il_5\, F^{\m \n}\,[\psib{}_\mu  \g_\n \,\g.\psi-\psib.\g \,\g_\m \psi_\n]\, .
\label{nm}
\eea
Our conventions are: Metric
$\eta_{\mu\nu}={\rm diag}(-,+,+,+)$,  
($\mu,\nu,..=0,\!..,3$, $i,j,..=1,\!.,3$) ; 
Dirac matrices:
$\{\gamma^{\mu},\gamma^{\nu}\}=2\,\eta^{\mu\nu}$, 
$\gamma^\mu{}^\dagger=\gamma^0\gamma^\mu\gamma^0$;
$\gamma^{\mu_1\ldots\mu_n}\equiv\gamma^{[\mu_n}\cdots\gamma^{\mu_n]}$
(we always (anti)symmetrise with unit weight); 
$\gamma^5=-i\gamma^{0123}$,
$\gamma^{\mu\nu\rho\sigma}=i\gamma^5\epsilon^{\mu\nu\rho\sigma}$.
Contraction of
all indices of a tensor with Dirac matrices is denoted by a hat, e.g.,
$\wh F=F_{\mu \nu}\g^\mu \g^\nu$.
The operator in the minimal term of~\eqn{action}
\be
\D_\m\equiv D_\m+\frac{1}{2}\,m\,\g_\m\, , \qquad 
D_\mu \psi_\nu=\d_\mu\psi_\nu+ieA_\mu\psi_\nu
\ee
also incorporates the usual mass term
$m\,\psib{}_\m\g^{\m\n}\psi_\n$; it 
satisfies 
\be
[\D_\m,\D_\n]=ieF_{\m\n}+\frac{1}{2}\, m^2\,\g_{\m\n}\, ,\qquad
[\D_\m,\g_\n]=m\g_{\m\n}\, .
\ee
Note that $il_5\, F^{\m \n}\,[\psib{}_\mu  \g_\n \,\g.\psi-\psib.\g
\,\g_\m \psi_\n]$, being diagonal in a Majorana basis,
is non-vanishing even for an uncharged real field;
on-shell (where $\g.\psi=0$) 
it does not contribute at lowest perturbative orders and
we therefore drop it until the general causality analysis 
in section~\ref{causal}.

The set~\eqn{nm} represents the most general non-derivative,
hermitean, parity-even
couplings linear in the field strength $F_{\m\n}$.
That they must constitute a 
five parameter family can also be seen upon
expanding the most general possible
$\psib{}_\m \Gamma^{\m\n\r\s}\psi_\n F_{\r\s}$
in a Fierz basis: There is a single coupling to 
$\gamma^{\m\n\r\s}$, three possibilities for $\g^{\m\n}$ and a single 
scalar $1\!\!1$ coupling. Thus the $\g^5 \wt F^{\m\n}$ of supergravity
($\wt F^{\m\n}\equiv(1/2)\,\e\tid^{\m\n\r\s}F_{\r\s}$) 
may be cast in the above basis as
\bea
\psib{}_\mu i\g^5 \widetilde F^{\mu\nu} \psi_\nu&=&
\psib{}_\mu F^{\mu \nu}\,\psi_\nu
\,-\,
\frac{1}{2}\,\psib{}_\mu \wh F \,\psi^\mu\nn\\&&
\,-\,
F^{\m \n }\,[\psib{}_\mu \g_\n \,\g.\psi+\psib.\g \,\g_\m \psi_\n]
\,+\,\frac{1}{2}\,\psib.\g\, \wh F\, \g.\psi\, .
\label{identity}
\eea

Two of the five parameters in~(\ref{action},\ref{nm}) may be eliminated by
requiring that the model describe  the correct DOF or, equivalently,
maintains the constraint count of the free theory:
The zeroth component of the equation of motion 
$R_0=\delta {\cal L}/\delta\psib\hspace{.5mm}{}^0$ 
involves no time derivatives and
is therefore a constraint eliminating four of the sixteen (complex) 
components of $\psi_\mu$. Another constraint eliminating 
four more components is still required before one can conclude that
on-shell 
half of the remaining components 
yield $2s+1=4$ physical DOF.
When $\psi_0$ appears linearly
in the action, as for the minimal theory, it is a Lagrange multiplier
imposing the
constraint $R_0=0$. 
Requiring its preservation under
time evolution, $\dot R_0=0$, yields the necessary additional
constraint. If instead $\psi_0$ appears quadratically ({\it i.e.}, as
$\psi_0^\dagger M \psi_0$ for some matrix $M$)
the $R_0=0$ equation now determines $\psi_0$, and
requiring $\dot R_0=0$
yields an equation of motion for $\dot\psi_0$.
This choice describes too few constraints/too many (propagating) DOF,
as compared to the free field.

One non-minimal model respecting the DOF count is the truncation of
$N=2$ supergravity with the cosmological, curvature and four-Fermi
terms omitted\footnote{This truncation should not 
be confused with the supersymmetry-preserving 
anti-de Sitter-Poincar\'e contraction,
mapping the model of~\cite{Freedman:1977aw} to the original $N=2$
model~\cite{Ferrara:1976fu}. The latter has a flat gravitational
background and only non-minimal, uncharged, Maxwell couplings.}.
It corresponds by~\eqn{identity} to the 
choice of parameters $l_1=-2$, $l_2=1/2$, $l_3=1$, 
$l_4=-1/2$ (and $l_5=0$) reproducing $\F_{\m\n}=-(F_{\m\n}+i\g^5\wt F_{\m\n})$.
[If however, as in~\cite{Ferrara:1992yc}, the further truncation 
excluding the gamma-trace components of $\psi_\mu$ in the
non-minimal sector is made, the DOF count is violated. This fact would seem to
make moot the causality claim there.]

Henceforth, we retain only models
linear in $\psi_0$, which, as is easily seen, is equivalent to demanding
$\F_{\m\n}=-\F_{\n\m}$. The corresponding relations amongst parameters are
\be
l_2+l_4=0\,,\qquad l_3+2l_4=0\, ,
\label{DOF}
\ee
and the non-minimal interactions reduce to the two combinations 
(dropping~$l_5$)
\be
\psib{}_\m\,\F^{\m\n}\psi_\n\,= 
\,[l_1+2l_2]\,
\psib{}_\mu F^{\mu\nu}\psi_\nu
\,-\,
2l_2\,
\psib{}_\mu \,i\g^5\wt 
F^{\m\n} \,\psi_\n \, .
\label{nm1}
\ee
We now consider the properties of these physical spin~3/2 theories.

\section{Low Energy Theorems}
\label{low}

LETs characterise soft photon scattering amplitudes
in terms of the mass, charge and magnetic
moment of the target~\cite{Low:1954kd,Lapidus:1961,Weinberg:1970bu}, 
independent of its internal structure,
relying only on gauge and Lorentz invariance plus
low photon frequency.
In this sense, LETs are purely kinematical  
and, irrespective of any causal pathologies, perturbative scattering
amplitudes formulated in terms of free asymptotic fields are
guaranteed to satisfy them. Therefore
LETs provide a simple way to map our 
parameters to the physical ones. 

We first study the vertex for the emission of a single low frequency
photon, then Compton scattering with small incoming (and outgoing)
photon frequencies. In each case the relevant LETs are
usually 
stated for stationary targets and Lorentz invariance 
is not manifest. 
So we first enunciate the dictionary with our relativistic Feynman
tree amplitudes.
A soft photon  
is invariantly defined by the requirement that
\be
\frac{\,\omega^a}{m}\equiv-\,\frac{p.k^a}{m^2} <\!\!< 1\, , 
\ee
for photon a's four-momentum $k^a_\mu$ and target $p_\m$.
In the laboratory frame $p_\m=(m,0,0,0)$, $\w^a$ reduces to
the usual photon frequency.

For each photon polarisation
$\varepsilon_\mu^a$ we employ Feynman gauge $k^a.\varepsilon^a=0$
along with the residual gauge fixing condition $p.\varepsilon^a=0$
so that in the lab.~frame one has
$\varepsilon_\mu^a=(0,\vec{\epsilon}\,{}^a)$ and $\vec k^a\cdot
\vec\epsilon{}\,^a=0$. We also utilise a covariant notation for the
target polarisations $u_\mu=u_\mu(p)$ and $\ol u'_\m=\ol u_\m(p')$ where
$p'_\m$ is the outgoing momentum of the target particle and the usual 
asymptotic on-shell conditions hold
\be
p.u=\g.u=0=(i\sl{p\tid} +m)u_\m\, ,\qquad \ol u'.p'=
\ol u'.\g=0=\ol u'_\m(i\sl{p\tid} '+m)\, .
\label{clam}
\ee
An explicit
representation for the spin~3/2
polarisations in terms of the usual massive spin 1 and spin 1/2 polarisations
$\varepsilon_\mu^l$ ($l=-1,0,1$) and $u^s$ ($s=-1/2,1/2$), respectively, 
is given by $u_\mu^{\pm 3/2}=\varepsilon_\m^{\pm 1} u^{\pm 1/2}$ and
$u_\m^{\pm 1/2}=$ $(\varepsilon_\mu^{\pm 1} u^{\mp 1/2}+\surd\!\tid2 \,
\varepsilon_\mu^0 \,u^{\pm 1/2})/\surd\!\tid3$~\cite{Auvil:1966}. Obviously,
inserting $u_\r=u_\r^\l$ and 
$\ol u'{}\!_\r=\ol u{}_\r^{\l'}$ with
$\l,\l'=-3/2,1/2,1/2,3/2$ in~\eqn{gen} below, $\M^{\m\n}_{\l'\l}$
(regarded as a matrix in the labels $\l$ and $\l'$) 
is a spin~3/2 irreducible representation of the Lorentz algebra.

The total spin matrix $\vec{S}$ is, as usual, the dual of the (spatial) 
Lorentz generators, $S^i=\frac i2\epsilon^{ijk}M_{jk}$, and the 
Lorentz generators act on the relativistic vector-spinor on-shell 
representation of the spin~3/2 polarisations $u_\rho$ according to
\be
M^{\mu\nu,\rho}{}_\sigma=\frac{1}{2}\, \g^{\m\n}\delta^\rho{}_\sigma
\,+\,
2\,\delta^{[\m|\rho|}\delta^{\n]}{}_\sigma\, 
\label{Lorentz}
\ee
so that 
\be\delta_{\rm Lorentz}u_\rho=
\frac12\,\lambda_{\mu\nu} \,M^{\mu \nu,\rho}{}_\s\,u^\s=
\lambda_\rho{}^\sigma u_\sigma+\frac14\,\wh \lambda\, u_\rho.
\ee
It is useful to define
\be
{\cal M}^{\m\n}\equiv \ol u'{}\!_\r\, M^{\m\n,\r}{}_\s\, u^\s\,
=\,\frac{1}{2}\,\ol u'{}\!_\r \g^{\m\n}\,u^\r+2\,\ol u'{}^{[\m}\,u^{\n]}
\, ,\quad
{\cal S}^i\equiv\frac{i}{2}\,\epsilon^{ijk}\,
\ol u'{}\!_\r\, M_{jk}{}^\r{}_\s\, u^\s\, .
\label{gen}
\ee

The LET for the  photon
vertex states that the amplitude for emission of a soft photon
by a stationary mass $m$, spin $s$  ``target'' 
is
\be
T_{fi}=
-\,\frac{i\mu}{s}\,(\vec{\epsilon}\times\vec{k})\cdot\vec{\cal
S}+{\cal O}(\omega^2)\, ;
\label{amp3}
\ee
transparent derivations of~\eqn{amp3} and~\eqn{LET} below
may be found in~\cite{Weinberg:1970bu}.
The  magnetic moment $\mu$ appearing in~\eqn{amp3} is related to the
charge/mass ratio of a spin $s$ particle
by the gyromagnetic ratio $g$, defined
by
\be
\mu\equiv\frac{egs}{2m}\, .
\ee 

The standard LET for Compton scattering 
reads
\bea
T_{fi}&=&-\,\frac{e^2}{m} \,\, \vec\e\,'\cdot\vec\e \;\;\ol u'. u 
\,+\,
\frac{ie\,\w}{m}\,\Big(\frac{2\mu}{s}-\frac{e}{m}\Big)\,
(\vec\e\,'\times\vec\e)\cdot \vec{\cal S}
\nn\\&&
-\,\frac{ie\,\mu}{\omega\,s}\Big(\vec\e\cdot\vec k\,'\,(\vec\e\,'\times\vec k\,')
-\vec\e\, '\cdot\vec k\,(\vec\e\times\vec k)
\Big)\cdot\vec{\cal S}\nn\\&&
-\,\frac{i\,\mu^2}{\omega\,s^2}\Big((\vec\e\,'\times\vec k\,')
\times(\vec\e\times\vec k)\Big)\cdot\vec{\cal S}+{\cal O}(\omega^2)\, .
\label{LET}
\eea
Our task now is to derive  
the amplitudes~\eqn{amp3} and~\eqn{LET}
in a Lagrangian framework and thereby 
relate the parameters of~(\ref{action},\ref{nm1})
to the physical ones which, (apart from $e$ and $m$) means the single
number $g$.

\subsection{The Soft Photon Vertex}

\label{softgamma}

The amplitude~\eqn{amp3} may, using the on-shell
conditions
for the target and soft photon polarisations, be expressed
in the manifestly Lorentz invariant form
\be
T_{fi}=\frac{i\mu}{2s}\;F_{\m\n}\,\M^{\m\n}+
{\cal O}(\omega^2)\, , \qquad F_{\m\n}=i(k_\mu \e_\nu-k_\nu \e_\mu)\, .
\label{amp3rel}
\ee
On-shell the interaction Lagrangian, including 
minimal and non-minimal couplings, becomes
\be
i{\cal L}^{\rm int}\bracevert_{\rm on-shell}=i\,T_{fi}=
e\,\ol u_\rho\, \epsilon\!\!/ \, u^\rho
\,+\,
\frac{e}{m}\,\Big[\,
l_1\,\ol u{}_\mu\,F^{\m\n}\,u_\nu
\,+\,
l_2\,\ol u_\rho\, \wh F\, u^\rho
\Big]\, .
\ee
The asymptotic Rarita--Schwinger equation 
may be used to derive the obvious generalisation
\bea
\ol u'{}\!_\rho (p') \g^\mu u_\sigma(p)&=&-\frac {i}{2m} \,(p+p')^\mu\,\,\ol
u'{}\!_\rho u_\sigma
+\frac {i}{2m} (p'-p)_\nu \,\ol u'{}\!_\rho
\g^{\mu\nu} u_\sigma 
\label{gordon1}
\eea
of the spin 1/2 Gordon identity.
Thus the amplitude derived from the non-minimal Lagrangian reads
\be
T_{fi}=
-\frac{ie}{2m}\,\Big(
2l_1\,\ol u{}_\m\,F^{\m\n}\,u_\n+\frac{1}{2}(4l_2-1)\,\ol u{}_\r\,\wh
F\,u^\r\, 
\Big)\, .
\label{SRV}
\ee
At this juncture, the amplitude seems quite different from that of a pure
$s=1/2$ system for which there is a one parameter family of $(g-2)$ values
read off from the $\wh F$ term with, of course, no counterpart to the
$l_1$ term. If the latter is to augment the $\wh F$ term to 
a coupling involving the full Lorentz generators as in~\eqn{gen}, the relevant
coefficient between the two terms must be $1/4$. There must be,
therefore, an identity relating these two terms for the LET to hold
and this is indeed the case (see~\eqn{shell} below). 
Alternatively, it is possible, for the choice of parameters
$l_1=-g/2=4l_2-1$ to directly satisfy the LET,
in which case the Lagrangian then reads
\bea
{\cal L}&=&{\cal L}_{\rm Min}+\frac{ie}{m}\,\psib{}_\m F^{\m\n}\psi_\n
\,+\,\frac{ie(g-2)}{4m}\,F_{\m\n}\,\psib_\rho
M^{\m\n,\r}{}_\s\,\psi^\s\nn\\&&
+\,\frac{ie(g-2)}{4m}\,
\Big(
F^{\m \n}\,[\psib{}_\mu  \g_\n \,\g.\psi+\psib.\g \,\g_\m \psi_\n]
\,-\,
\frac{1}{2}\,\psib.\g\, \wh F\, \g.\psi
\Big)\, .
\label{theory}
\eea
Together, the first two terms of ${\cal L}$
produce a $g=2$ coupling since the minimal
Lagrangian gives the spin~1/2 part of the Lorentz generators via
the Gordon identity above and the non-minimal coupling to $F^{\m\n}$
yields the spin~1 contribution (exactly the same term
required for a spin~1 vector boson to have  $g=2$).
The third term is a direct coupling to the total Lorentz generators 
and yields an anomalous magnetic moment coupling, the remaining ones
being required to ensure the correct DOF in the 
$g\neq 2$ case. One might speculate whether this
theory may be physically more desirable for  higher order processes
not completely determined by LETs.
It is interesting to further note that it does not 
include truncated $N=2$ supergravity.

Since LETs are just a statement about parts of
amplitudes determined completely by kinematics, they  
should be reproduced for any choice of 
parameters \hspace{-.1cm}\footnote{In~\cite{Freedman:1977aw} the Thompson
limit for Compton scattering was obtained in the truncated $N=2$
supergravity model, a calculation that seemed to hinge on
delicate cancellations due to 
the particular form of the supergravity non-minimal couplings.
However, since the Thompson limit is dictated
by the lowest order LET,
our generic computation always guarantees this result.}.
The on-shell identity 
\be
\ol u'_\rho(p')\,\g_{\m\n}\,u^\rho(p)=2\,\ol u'_{[\mu}(p')\,u_{\n]}(p)+
{\cal O}(\w)
\label{shell}
\ee
is easily verified in the frame $p_\m=(m,0,0,0)$. 
This low energy equality states
that on-shell the spin~1/2
and spin~1 parts of the Lorentz generators may be traded
against one another. Hence the low energy result is
precisely reproduced by {\it any} of the 
non-minimal Lagrangians, the gyromagnetic ratio
being
\be
g=\frac{2}{3}-\frac{4}{3}\,(l_1+2l_2)\, ,
\label{g}
\ee
a sum of minimal and non-minimal contributions.
The model in~\eqn{theory} is clearly a
sub-case of this result.
For minimal coupling the well known result $g=1/s=2/3$
emerges~\cite{Belinfante:1953}. 
Observe, that the parameters, $l_1=-2$ and $l_2=1/2$, of truncated 
$N=2$ supergravity
also yield 
$g=2$~~\cite{Ferrara:1992yc}.   
Finally, we note, in passing, that in the basis~\eqn{nm1} only the $F_{\m\n}$
coupling contributes to the gyromagnetic ratio~\eqn{g},
independent of the $i\g^5 \wt F_{\m\n}$ term. This is not
surprising since $\g^5$ mixes the ``large'' and ``small'' components of the
vector-spinor $u_\m$ defined by projection with 
respect to $i\g^0$,
and is higher order in the soft
photon expansion. This term will contribute to higher quadrupole and
octupole moments. The extension of our work to such moments
is an interesting but separate issue.

\subsection{Compton Scattering}

A useful check on our vertex result~\eqn{g} is to compute
the amplitude for Compton scattering.
In particular, since the gyromagnetic ratio must be precisely $g=2$
for the optical theorem to hold~\cite{Weinberg:1970bu},
the latter will produce
an additional relation between the parameters $l_1$ and $l_2$.

Let us denote incoming and outgoing
photon momenta by $k_\m$ and $k'_\m$, respectively; 
the mass shell condition
$p'{}^2=-m^2$ implies that the difference $\w'-\w=k.k'/m$ is second
order in this expansion. Therefore we 
eliminate $\w'$ (using this relation) and $p'$  (by momentum
conservation);
one can then evaluate the order of
any expression simply by counting the number of four-vectors $k$ and
$k'$ in it.
We now evaluate~\cite{Vermaseren:1991} 
the relevant $s$ and $u$ channel diagrams
of Figure~1 in this limit, 
using the vertices of~\eqn{action} and the free propagator 
\be
S^F_{\m\n}(p)=
\frac{-i}{p^2+m^2}
\Big[
(\eta_{\mu\nu}+\frac{p_\mu p_\nu}{m^2})(i\sl{p\tid} -m)+
\frac{1}{3}(\frac{ip_\mu}{m}-\gamma_\mu)
(i\sl{p\tid} +m)(\frac{ip_\nu}{m}-\gamma_\nu)
\Big]\, .
\label{prop}
\ee
To extract the leading and next-to-leading terms in the low energy
expansion of the
amplitude, we need the Gordon identity~\eqn{gordon1}
of the previous section along with the following generalisation,
\bea
\ol u _\lambda'(p') \g^{\mu\nu\rho} u_\kappa(p)
=-\frac{3i}{2m} \,(p+p')^{[\mu}\,\,\ol u_\lambda'
\g^{\nu\rho]}u_\kappa+\frac {i}{2m} \,\ol u_\lambda'
\g^{\mu\nu\rho\sigma} u_\kappa \,\,(p'-p)_\sigma\, .
\label{gordon2}
\eea
Note that in the lab.~frame, the 
``Dirac'' equation~\eqn{clam} reads
$\gamma^0 u=imu$ 
and using the above Gordon
identities, expressions 
such as $\ol
u'_\rho \gamma^{\mu\nu}\veps_\mu k_\nu u_\sigma$ are effectively
equal to $\ol u'_\rho
\gamma^{ij}\e_i k_j u_\sigma$ at leading order in the low energy expansion.
As a simple gauge invariance check, before imposing the
residual gauge choice $p.\veps=0=p.\veps'$ and taking the low energy
limit, we verified that our amplitude satisfies transversality in each
photon line separately.

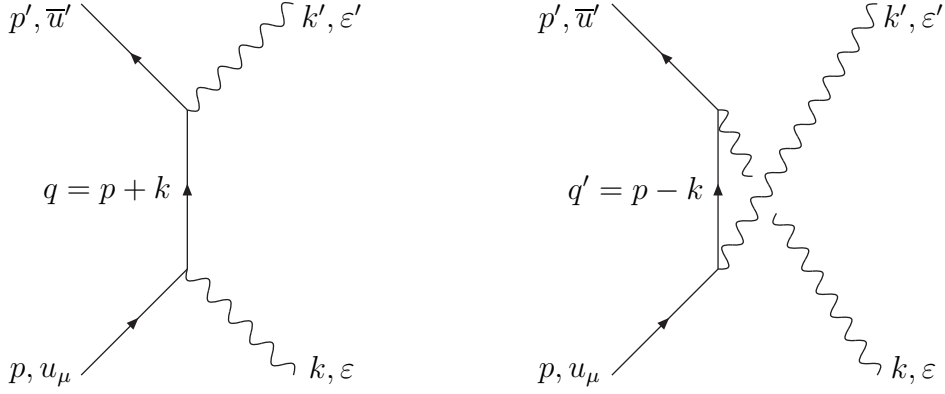
\begin{figure}
\begin{center}
\begin{picture}(400,160)(0,0)
\ArrowLine(50,10)(90,50)
\ArrowLine(90,50)(90,110)
\ArrowLine(90,110)(50,150)
\Photon(130,10)(90,50){-3}{5}
\Photon(90,110)(130,150){-3}{5}
\Text(35,10)[]{$p,u_\mu$}
\Text(35,145)[]{$p',\ol u'$}
\Text(60,80)[]{$q=p+k$}
\Text(145,12)[]{$k,\veps$}
\Text(365,12)[]{$k,\veps$}

\Text(145,145)[]{$k',\veps'$}
\Text(365,145)[]{$k',\veps'$}

\ArrowLine(250,10)(290,50)
\ArrowLine(290,50)(290,110)
\ArrowLine(290,110)(250,150)
\Photon(350,10)(312,71){-3}{7}
\Photon(303,85)(290,110){3}{3}
\Photon(290,50)(350,150){-3}{11}
\Text(235,10)[]{$p,u_\mu$}
\Text(235,145)[]{$p',\ol u'$}
\Text(260,80)[]{$q'=p-k$}
\end{picture}
\end{center}
\vspace{-.4cm}
\label{thefigure}
\caption{Compton scattering kinematics}
\end{figure}

Once again, applying only the Gordon 
identities~\eqn{gordon1} and~\eqn{gordon2},
the amplitude satisfies the LET only for the choice of
parameters  of the Lagrangian~\eqn{theory}. However, once one applies
further the on-shell identity~\eqn{shell} along with the additional identity
(equivalent to~\eqn{shell} upon contracting indices $\m$ and $\s$)
\be
\ol u'{}_{[\m}\g_{\n][\r}u_{\s]}=\ol u'{}_{[\r}\g_{\s][\m}u_{\n]}+
{\cal O}(\w)
\ee 
one finds for the amplitude (in an obvious 
matrix notation for vector indices)
\bea
T_{fi}&=&-\,\frac{e^2}{m}\;\ol u'.u\,\veps'.\veps
\,+\,\frac{e^2}{4m^3}\,(g-2)^2\,p.k\,(\veps'.\M.\veps)
\nn\\&&
-\,\frac{ie^2\,g}{4m\,p.k}\,[{\rm tr}\,(F'.\M)\,\veps.k'
-{\rm tr}\,(F.\M)\,\veps'.k]
\nn\\&&
+\,\frac{e^2g^2}{4m\,p.k}
\,{\rm tr}\,(F'.\M.F)\,+\,{\cal O}(\w^2)
\label{comrel}
\eea
where, as in the previous section $g$ stands for the combination~\eqn{g} of
parameters $g=2/3-4/3\,(l_1+2l_2)$.
It is not difficult to verify that in 
the lab.~frame, the amplitude~\eqn{comrel} precisely reproduces
the LET~\eqn{LET}.

\subsection{Unitarity and $g=2$}

We close this section with some comments on unitarity. As mentioned,
low energy unitarity 
imposes the value $g=2$
for any spin~\cite{Weinberg:1970bu}. The idea is that, 
(for a particle interacting
electromagnetically only), the
optical theorem constrains the low energy forward ($k=k'$) limit
of the scattering amplitude~\eqn{LET}
\be
T_{fi}=-\frac{e^2}{m} \,\, \vec\e\,'\cdot\vec\e \;\;\ol u'. u \,-\,
\frac{ie^2\,\w}{4m^2}\,(g-2)^2\,(\vec\e\,'\times\vec\e)\cdot \vec{\cal S}
\,+\,{\cal O}(\w^2)
\ee
to have no contribution linear in $\w$, thus requiring $g=2$.
Clearly, if one wishes to apply this criterion to our model,
one can simply take the choice of parameters $l_1+2l_2=-1$.
Of course, in reality, one may be interested in
an effective description of a composite particle participating
in the strong interactions with $g\neq2$.
Here the models 
with  general values of $g$ are suitable.

The quite distinct criterion of  tree 
unitarity~\cite{Cornwall:1974km,Piccinini:1984dd,Ferrara:1992yc}
concerns the high energy behaviour of the theory. Partial wave
amplitudes are subject to (constant) 
unitarity bounds which may, in principle, be violated by 
tree level amplitudes. For effective
theories these bounds determine the energy scale at which the
effective description fails and new physics enters 
(e.g., inapplicability of the  
Fermi weak interaction theory  
beyond $m_{W}$). This argument does not, however, rule
out the low energy applicability of the effective theory.
The failure of tree unitarity in the minimal model
was first observed in~\cite{Piccinini:1984dd}. [Their inference
of a connection between this and acausality seems unwarranted,
however, given that the $g=2$ tree
unitary model also fails to propagate causally.] 
Let us, nonetheless, review the tree level unitarity 
argument~\cite{Piccinini:1984dd,Ferrara:1992yc} in more detail.
Massive higher spin propagators such as~\eqn{prop}
contain inverse powers of the mass.
In tree level Green functions, for general kinematical configurations 
where all momenta are large,
these terms lead to contributions growing with positive powers
of the energy $E$. In addition to being  dangerous
for renormalisability when higher order loops are constructed from
trees,
they eventually violate
partial wave unitarity bounds.
There is, however,  
a quite general mechanism related to gauge invariance 
to remove this
undesirable high energy behaviour. 
Namely, if one investigates the
worst powers of inverse mass in the propagator~\eqn{prop}
\be
\frac{\, \frac{\ts 2}{\ts 3}\,\frac{\ts p_\m\sl{p\tid}  p_\n}{\ts m^2}\,}
{p^2+m^2}
\ee
one see that the operators $p_\m$ and $p_\n$ generate a linearised
gauge transformation (in our case a linearised local supersymmetry
transformation) 
\be
\delta \psi_\m=\d_\m\veps
\ee
at the vertices to which the propagator is attached. Hence requiring
the on-shell vertices to satisfy the appropriate supersymmetric Ward
identity will lead to cancellation of $1/m^2$ contributions. Obviously
one can apply this procedure to further constrain the 
non-minimal couplings.
The supergravity solution with
$l_1=-2$ and $l_2=1/2$ uniquely satisfies this criterion via the usual
supersymmetry Ward identity (the parameters
$l_3$ and $l_4$ remain undetermined since $\g.\psi=0$ on an external line)
and in this sense the supergravity-inspired model may be favoured
amongst possible models although, as argued above, 
at the more phenomenological level of effective theories 
this choice is not particularly compelling.

\section{Causality and Quantisation}
\label{causal}

As we shall discuss, a study of 
causality\footnote{The causality study for various couplings
to external fields in~\cite{Hagen:1982ez} did not include non-minimal 
couplings; none of the models considered there was causal either.}
amounts to investigating whether the constraints, required to ensure the
correct physical DOF, are consistent.
In particular, one may find that, for some critical value of
the external field $F_{\m\n}$, the secondary Lagrangian
constraint\footnote{Our terminology is as follows: 
For a first order system devoid of gauge
invariances, the primary (Lagrangian) constraints are simply
any field equations without  time
derivatives. Requiring that the primary constraints are
preserved by time evolution leads to secondary  constraints 
and so forth. Precisely the same constraints arise as second class
secondary and tertiary constraints, respectively, in a canonical Dirac
analysis~\cite{Dirac}.} 
may no longer be inverted to solve for the Lagrange multiplier variable
$\psi\hspace{.05mm}_0$. From a canonical viewpoint, this implies that the
Dirac bracket governing dynamics 
on the constraint surface is ill-defined at this
point~\cite{Johnson:1961vt} and yields a pathology that, of course, extends
to the corresponding quantum mechanical
canonical commutators. In terms of the field equations, this
pathology implies that the model permits superluminal
propagation~\cite{Velo:1969bt}. A brief review of the precise
relation between causality breakdown and consistency of constraints
is given at the end of this section.

The key point is to find, and study the consistency of, eight (complex)
constraints amongst the sixteen field components $\psi_\mu$;
the equations of motion then reduce these to
four physical DOF.

The field equation 
derived from~\eqn{action} is
\be
R_\mu\equiv
{}^{\textstyle \delta S}\!\!\Big/\!\!_{\textstyle \delta \psib\,^\mu}
=\g_{\m\n\r}\,\D^\n\psi^\r+\frac{ie}{m}\,\F_{\m\n}\,\psi^\n=0\, .
\label{eom}
\ee
Since $R_0$
does not involve time derivatives of any fields
\be
\Theta_1=\g^0 R_0=
\g_{ij}\,\D^i\psi^j+\frac{ie}{m}\,\g_0\,\F^{0i}\,\psi_i\,=0 \, .
\label{primary}
\ee
is a primary constraint.
As explained in section~\ref{themodel}, 
a correct
DOF count requires that $\F_{00}=0$ so that~\eqn{primary} does
not determine the Lagrange multiplier $\psi_0$.

Before taking the divergence of $R_\mu$ to determine the secondary
constraint we employ the relation (equivalent to~\eqn{primary}
on-shell)
\be
\g.R=2\,(\Sl \D-3m)\g.\psi-2\,\D.\psi+\frac{ie}{m}\,\g.\F.\psi=0
\label{geom}
\ee
to rewrite the field equations $R_\mu$ as
\bea
R_\m&=&(\Sl \D-m)\psi_\m-(\D_\m-\g_\m\,[\Sl \D-2m])\,\g.\psi-
\g_\m\,\D.\psi
+\frac{ie}{m}\,\F_{\m \n}\,\psi^\n\nn\\
&=&(\Sl \D-m)\psi_\m-(\D_\m-m\g_\m)\,\g.\psi+\frac{ie}{2m}\,
\g_\n\g_\m\F^{\n\r}\psi_\r\;=\;0\, .
\label{neom}
\eea
In particular, in temporal gauge $A_0=0$, 
the equations of motion for the
spatial components of the Rarita--Schwinger field are
\be
\g_0 R_i=\dot\psi_i+\g_0\,(\vec\g\cdot\vec\D-\frac{1}{2}\,m)\,\psi_i
-\D_i\,\g_0\g.\psi
+\frac{ie}{2m}\,\g_0\,\g_\n \g_i \F^{\n\r}\psi_\r=0\, .
\label{dot}
\ee
We now obtain a secondary Lagrangian constraint from
\bea
\Theta_2\,=\,\D.R&=&
-\frac{3}{2}\,m^2\,\g.\psi
+\frac{ie}{2}\g_{\m\n\r}\,F^{\m\n}\psi^\r
+\frac{ie}{m}\,\D.\F.\psi\nn\\
&=&
-\frac{3}{2}\,m^2\,\g.\psi
+\frac{ie}{2}\g_{\m\n\r}\,F^{\m\n}\psi^\r
+\frac{ie}{m}\,(\D^i\F_{i\nu}+\frac{1}{2}\,m\g^0\F_{0\n})\,\psi^\n
\nn\\&&
+\frac{ie}{m}\,(\dot\F\tid^{0i}\,\psi_i+\F\tid^{0i}\,\dot\psi_i)\;=\;0\, ;
\label{second}
\eea
since $\dot\psi_i$ may be eliminated via~\eqn{dot}, $\Theta_2$
constitutes a second
independent algebraic relation amongst field components. [Again,
observe that~\eqn{second} would contain a term 
$(ie/m)\,\F_{00}\,\dot\psi_0\hspace{.05mm}$ for $\F_{00}\neq 0$ 
and become an equation of motion rather than a constraint.]

Upon substituting~\eqn{dot} into~\eqn{second}
we concentrate on the coefficient matrix of the Lagrange multiplier $\psi_0$ 
in $\Theta_2$, since $\psi_0$ 
must be determined by this relation:
\bea
\Theta_2&\equiv& \g^0 R\; \psi_0+\cdots 
\nonumber
\eea
\be
R=-\frac{3}{2}\, m^2+\frac{ie}{2}\,\g_i F^{ij} \g_j
-\frac{ie}{m}\,\g_0\,[\D_i,\F\tid^{0i}]
-\frac{e^2}{2m^2}\,\g_0\F\tid^{0i}\g_0\,\g_j\g_i\,\F\tid^{0j}\, .
\label{test}
\ee
In terms of the electric and magnetic fields 
($E^i=F^{0i}$, $B^i=\wt F^{0i}$)
\be
\F^{0i}=l_1\,E^i+(2l_2\,\delta^{ij}+
l_5\,\epsilon^{ijk}\g_0\g^5\g_k)\,(E_j-i\g^5 B_j)
\ee
the critical matrix $R$ whose loss of invertibility would leave 
$\psi_0$ (partly) undetermined, is
\bea
R&=&
-\,\frac{3}{2}\,m^2
-e\,[1-2l_2]\,\g_0\g^5\vec\g\cdot\vec B\
-2e\,l_5\,\g_0\vec \g\cdot\vec E
\nn\\&&
+\,\frac{e^2}{2m^2}([l_1+2l_2]^2\,\vec E\,^2+[2l_2]^2\,\vec B\,^2\,)
\nn\\&&
+\,\frac{e^2}{m^2}\,[2l_2(l_1+2l_2)+2l_5^2]\,\g_0\,\vec\g\cdot(\vec E\times\vec
B)
+\frac{2e^2}{m^2}\,l_1l_5\,\g^5\vec E\cdot \vec B
\nn\\&&
-\,\frac{ie}{m}\,[l_1+2l_2]\,\g_0\,\vec\nabla\cdot\vec E
+\,\frac{ie}{m}\,l_5\,\g^5\vec \g\cdot(\vec \nabla\times\vec E
-i\g^5\vec \nabla\times\vec B)
\, .
\label{are}
\eea
A pathology in quantisation and causality of the model thus occurs whenever
$\det R=0$ 
as a function of the
background fields. To see how this occurs consider first a  
pure constant electric background. The determinant 
obviously develops a zero
for a critical value of the electric field (with the choice $l_5=0$)
\be
\vec E\,^2=3\,\left(\frac{m^2}{e[l_1+2l_2]}\right)^2\, .
\ee
But the choice of parameters $l_1+2l_2=0$ and $l_5$ arbitrary,
cannot yield a causal model either, because the determinant vanishes 
in a pure magnetic background whenever
\be
-\frac{3}{2}\,m^2+\frac{e^2}{2m^2}\,[2l_2]^2\vec{B}^2=\pm\,
e\,[1-2l_2]\,|\vec B|\, .\label{bound}
\ee
Truncated supergravity ($l_2=1/2$, $l_5=0$) and minimal coupling 
($l_1=l_2=l_5=0$)
have critical
field values $\vec B^2=3m^4/e^2$ and $\vec B^2=(3m^2/2e)^2$,
respectively (the latter being the well known result
of~\cite{Johnson:1961vt,Velo:1969bt}). 
In the general case~\eqn{bound}, has a solution
whenever the quadratic
\be
P(\beta)\equiv (2 l_2\,\beta)^2\pm2\,(1-2l_2)\,\beta-3=0\, ,\qquad
\beta \equiv |e\vec B|/m^2\, ,
\label{polly}
\ee
has a solution for $\beta>0$. 
Clearly, for any non-zero value of $l_2$, $P(\beta)$ is
positive for large enough $\beta$ and negative near $\beta=0$
so it  always has a zero for some positive $\beta$:
{\it All} models, minimal or non-minimal, exhibit
pathological behaviour\footnote{Precisely the same analysis for a pure
electric field yields the same result. Also even an
uncharged (real) Majorana field, with only $l_5\neq0$,
displays acausal propagation.}.

It is interesting to speculate whether further non-minimal couplings
may restore causality. In particular, the choice $l_2=1/2$, $l_5=0$ at least
removes the terms linear in $\vec B$ and $\vec E$
in~\eqn{are} responsible for the
original pathology of the minimal model. [In this respect, we note that
this choice along with $l_1=-2$ is that of supergravity for which
the field-dependent terms of~\eqn{are}
are proportional to the electromagnetic energy
density, Poynting vector and charge density and for this reason 
causality is preserved there, upon taking gravity and the
electromagnetic field dynamical.]
In Appendix~\ref{appB} we generalise~\eqn{are} to
arbitrary non-minimal couplings and show that broad classes of
couplings fail to propagate causally.

Finally, as promised, we briefly review the argument linking the
appearance of zeroes in $\det R$ to acausal propagation
in the field equations.
The computation of~\cite{Velo:1969bt} amounts to studying the Cauchy
problem of~\eqn{eom} and solving for the characteristic surfaces
that determine the maximal speed of propagation\footnote{In more
physical terms, this is akin to solving the equations of motion in a
high energy eikonal limit $\psi_\m=\Psi_\m \,\exp(i\tid t\tid x.\xi)$ with
$t\rightarrow\infty$. Clearly, timelike solutions for $\xi_\m$ indicate
superluminal propagation~\cite{Ferrara:1992yc}.}. This is simply
achieved by recalling that characteristics are
determined by discontinuities of the highest order derivative terms
appearing in the equations of motion~\cite{Madore:1973}. If we denote
the discontinuity $[..]$ of the first derivative of the Rarita--Schwinger
field across the characteristic as
\be
[\partial_\m\psi_\nu]=\xi_\m \Psi_\n
\ee
where $\Psi_\n$ is a non-zero vector-spinor field, 
then causal propagation forbids
timelike $\xi_\mu$. However from the field
equation~\eqn{eom} 
and its gamma-trace~\eqn{geom} we learn
\bea
[\R_\m-\frac{1}{2}\,\g_\m\g.R]&=&\sl \xi \Psi_\m-\xi_\m \g .\Psi\label{disa}\\
{}[\g.R]
&=&2(\sl \xi \g .\Psi-\xi .\Psi)
\label{disb}
\eea
and in turn
\be
\xi^2\Psi_\m=\xi_\m\xi.\Psi\, .
\label{disc}
\ee
Proceeding by contradiction we take $\xi_\m=(1,0,0,0)$ (timelike) 
without loss of generality since the original~\eqn{eom} is Lorentz covariant.
We now need only study the leading discontinuities in time derivatives
and in particular
\be
[\dot\Theta_2]=\g^0R\,\Psi_0=0\, .
\ee
which admits no non-vanishing solution for $\Psi_0$ unless $\det R=0$, 
the condition studied above.

\section{Discussion}
\label{conclusion}

We have seen that the most general
charged massive spin~3/2 theory with non-minimal couplings linear
in the electromagnetic field strength is described by the 
two parameter family
\bea
{\cal L}&=&-\,\psib\,^\mu \g_{\m\n\r}\,\D^\n\psi^\r
\,+\,\frac{ie}{m}\,\psib{}_\m\,F^{\m\n}\psi_\n\nn\\&&
+\,\frac{3ie}{4m}\,(g-2)\,\psib{}_\m\,F^{\m\n}\psi_\n
\,-\,\frac{2iel_2}{m}\,\psib{}_\m\,\g^5\wt F^{\m\n}\psi_\n\, ,
\label{result}
\eea
two of the other {\it a priori} admissible parameters being excluded by
DOF consistency; the third, corresponding to a diagonal Majorana 
coupling, did not affect  our low energy or causality results.
The physical interpretation of the first three terms in~\eqn{result} was
provided by studying LETs.
The first  is the usual minimally coupled 
Rarita--Schwinger theory with intrinsic gyromagnetic ratio $g=2/3$. Minimal
coupling for half integer systems yields  only the spin~1/2
contribution to the Lorentz generators,  while the second coupling is the 
spin~1 Pauli term required for $g=2$.
Although $g=2$ is required for low energy
unitarity of amplitudes
describing pure electromagnetic interactions, more general
phenomenological applications deal with the case $g\neq2$, 
and one may safely include the anomalous magnetic moment coupling given
in the third term. [Recall that at low energy, a coupling to $F^{\m\n}$ 
is equivalent to a coupling to the full Lorentz generators up to a
factor $1/3$, by virtue of the identity~\eqn{shell}.]

The fourth term is more subtle, as it does not contribute
at low energy until quadrupole order. 
It will be an exercise of some physical importance 
to relate $l_2$, as well as effective Lagrangians including gradients
of $F^{\m\n}$, to
multipole moments along the lines of the method presented here for
the magnetic dipole~\cite{Deser:2000}.
As we have noted, however,
there are already 
interesting values of $l_2$. The first, implying tree unitarity,
is $l_2=1/2$ as well as
$g=2$ and is a truncation of $N=2$ supergravity 
(along the lines of~\cite{Ferrara:1992yc}, but maintaining the
correct DOF).
Another possibility is to
take 
$l_2=-(g-2)/8$ which (see~\eqn{identity}) promotes the anomalous
magnetic moment coupling of the third term
from one involving only spin~1 to one already including
the total
Lorentz generators in the Lagrangian. 

Our study of 
causality showed that no model maintaining the correct
$\mbox{DOF}$ avoids sharing the pathology of the minimal one. 
In fact
this result applies to a very broad class of non-minimal couplings
(beyond just linear in the field strength);
the criteria described in section~\ref{causal}
determine the causality of any non-minimally coupled model.
An interesting issue under study is whether
including gravity dynamically can improve upon this situation; 
certainly for supergravities~\cite{Deser:1977uq} this is the case, 
although the minimal
model in curved space is known to still suffer the usual 
difficulties~\cite{Madore:1975}.

Finally, and perhaps most physically relevant,
the models we have studied, despite the 
formal causal pathologies of the
interacting fields, 
provide a useful parametrization for an  effective low energy
description of higher spin excitations:
They are a field theoretical framework for the generic LET properties.

\section*{Acknowledgements}

We are indebted to H.~Schnitzer and S.~Weinberg for reminding us of the
Wigner--Eckart theorem's universality and thank also 
M.~Porrati and P.~van Nieuwenhuizen for discussions. 
This work was supported by the
National Science Foundation under grant PHY99-73935 and by 
the Australian Research Council.

\begin{appendix}

\section{Appendix: Higher Spin Soft Photon Vertices}
\label{appA}

The results of section~\ref{softgamma} are easily generalised to
arbitrary higher spin targets. For higher integer spin $s$ we employ a
complex symmetric tensor field $\phi_{\m_1\cdots\m_s}$. On-shell, 
$\phi_{\m_1\cdots\m_s}$ is asymptotic to a free field satisfying
\be
(\Box-m^2)\,\phi_{\m_1\cdots\m_s}=0=\phi_\m{}^\m{}_{\m_3\cdots\m_s}=
\d_\m\,\phi^\m{}_{\m_2\cdots\m_s}\, . 
\label{as1}
\ee
The corresponding half integer spin $s\equiv n+\frac12$  
representation 
is a complex Dirac symmetric tensor-spinor $\psi_{\m_1\cdots\m_n}$
obeying free field equations
\be
(\sl\d+m)\,\psi_{\m_1\cdots\m_n}=0=\g_\m\psi^\m{}_{\m_2\cdots\m_n}=
\psi_\m{}^\m{}_{\m_3\cdots\m_n} \,
=\d_\m\psi^\m{}_{\m_2\cdots\m_n}
\, .
\label{as2}
\ee

Here too it is essential to determine 
appropriate non-minimal couplings to 
lower trace and gamma-trace field components
to ensure a correct DOF count in an electromagnetic background. 
Although we have not yet studied this generally, it is
tempting to assume that antisymmetry of the non-minimal
interactions is again the correct criterion.
In this Appendix however, we ignore these couplings since they are
irrelevant to the soft photon vertex.

The general Lagrangian is the sum of
the minimally coupled 
massive higher spin action\footnote{Note that
in~\cite{Singh:1974qz} the action is in terms of traceless and
gamma-traceless fields symmetric in vector indices along with auxiliary
fields corresponding to all possible traces and gamma-traces.
Using field redefinitions one may work, equivalently, with unconstrained
symmetric fields as above.} of~\cite{Singh:1974qz} plus the most general 
non-minimal couplings (ignoring
trace couplings) denoted by ${\cal L}_{\rm NM}$.
For integer spins\footnote{Here and throughout, we have
ignored derivative couplings, although they may also  
contribute to the magnetic moment; this is illustrated 
in~\cite{Schwinger:1970}.}, 
\be
{\cal L}_{\rm NM}=-\,ie\,l_1\,\phi^*_\m F^{\m\n}\phi_\n
\label{snooze}
\ee
and for half integer spins (the precise analogue of~\eqn{nm}),
\be
{\cal L}_{\rm NM}=-\,\frac{ie}{m}\,
\Big[ \,l_1\,\psib{}\!_\m F^{\m\n}\psi_\n+l_2\,\psib \,\wh F\,\psi\,
\Big]
\ee
in the terse notation, appropriate for bilinears, 
that drops any indices contracted directly
between a field and its complex conjugate (so that, for example,
$\psib \,\wh F\,\psi\equiv\psib{}\!_{\r_1\cdots\r_n}\,\wh
F\,\psi^{\r_1\cdots\r_n}$).

The LET for the emission of a single photon for arbitrary spin
target is
\be
T_{fi}=\frac{i\mu}{2s}\;F_{\m\n}\,\M^{\m\n}+
{\cal O}(\omega^2)
\label{ampagain}
\ee
where now the Lorentz generators in a higher spin representation are given
by
\be
\M_{\m\n}=2ms\,\varepsilon'^*_{[\m}\,\varepsilon_{\n]}
\ee
for integer spins, and 
\be
\M_{\m\n}=\frac12\,\ol u' \g_{\m\n}u+2s\,\ol u'_{[\m}\,u_{\n]}
\ee
for half integer spins,
where the initial, $\varepsilon_{\m_1\cdots\m_s}$
and $u_{\m_1\cdots\m_n}$, respectively, and final target
polarisations (with primes) 
satisfy the usual conditions implied by~\eqn{as1} and~\eqn{as2}.

We must now compute the on-shell vertices in the soft photon limit and
compare the results with the LET~\eqn{ampagain}. The integer spin case
is simple and we find a gyromagnetic ratio
\be
g=\frac{1}{s}-\frac{2l_1}{s}
\ee
where we have included a contribution $g=1/s$ from the 
minimally coupled Lagrangian. (Of course, unlike the half integer case, 
a second order system is well known to be ambiguous due to
possible partial integrations before minimal coupling.
In fact the minimal model alone can yield any gyromagnetic ratio
between $g=0$ and $g=1/s$, the case quoted above being attained
by writing the Lagrangian in first order form and only thereafter
coupling minimally~\cite{Singh:1974qz}.)

The half integer case directly follows
section~\ref{softgamma}; we must include the minimal
interaction
\be
{\cal L}^{\rm int}_{\rm Min}=-ie\,\psib\sl A \psi
\ee
to which the Gordon identity~\eqn{gordon1} may be applied unaltered.
Once again, there is a special model which requires no further low
energy identity identities to fulfill the LET, when $n l_1=4l_2-1$.
The LET is satisfied, of course, for all parameters $(l_1,l_2)$
once one derives the obvious generalisation of the
identity~\eqn{shell},
implying
\be
\M_{\m\n}=s\,\ol u'{} \g_{\m\n} u\,+\,{\cal O}(\w)\, .
\ee
As a result one finds
\be
g=\frac{1}{s}-\frac{2(l_1+2l_2)}{s}\, ,
\ee
which clearly reproduces~\eqn{g} for $s=3/2$. Note that the 
higher spin analogue
of~\eqn{nm1} again implies that an $i\g^5 \wt F_{\m\n}$ coupling does
not contribute at linear order in $\w$, so that low energy physics
is encapsulated by a single magnetic moment coupling.

\section{Causality of General Non-Minimal Models}

\label{appB}

The curious reader may wonder whether there exist causal
non-minimal couplings for more general
functions $\F^{\m\n}$ of
field strengths.
The most general antisymmetric (so that DOFs are maintained) one 
is
$\F^{\m\n}=W^{\m\n}+i\g^5 X^{\m\n}+i\g^{\m\n}Y+\g^5\g^{\m\n} Z$
where $W^{\m\n}$ and $Y$ are parity even, $X^{\m\n}$ and $Z$ are
parity odd, all built from $F_{\m\n}$, $\wt
F_{\m\n}$, $(F_{\r\s}F^{\r\s})$, $(F_{\r\s}\wt F^{\r\s})$
and field-gradient dependent terms. [For brevity we omit the
diagonal, Majorana, $l_5$ term.]
Causality is determined by substituting this expansion into the
matrix~\eqn{test} and searching for zeroes in its determinant.
If we set $Z=0$ (in any case $Z$ must be an odd function of the 
axial scalar $(F_{\r\s}\wt F^{\r\s})$, so vanishes for pure electric
or pure magnetic field configurations) then the analysis of this
determinant is almost identical to that performed above:
The third term $Y$ is a field-dependent mass term so
replacing the operator $\D_\m\rightarrow D_\m+\frac{1}{2}m_*\g_\m$
where $m_*=m+Y$, calling
$\F^{i0}=E_*^i+i\g^5 B_*^i$ ($E_*^i=W^{i0},B_*^i=X^{i0}$)
and derivatives of $m_*$ no longer vanish. The matrix $R$ of~\eqn{are}
then reads
\bea
R&=&-\frac{3}{2}\,m_*^2+\frac{e^2}{2m^2}(\vec E_*^2+
\vec B_*^2\,)
+e\,\g_0\g^5\vec\g\cdot(\vec B+\frac{m_*}{m}\,\vec B_*)\nn\\&&
+\frac{ie}{m}\,\g_0\,(\vec\nabla\cdot\vec E_*-i\g^5\vec\nabla\cdot\vec
B_*)
+\vec\g\cdot\vec\nabla m_*
-\frac{e^2}{m^2}\,\g_0\,\vec\g\cdot(\vec E_*\times\vec
B_*)\, .
\eea
For the (simplest) case $F_{\m\n}$ constant, $\det R$ vanishes whenever
\be
\Big[-\frac{3}{2}m_*^2+\frac{e^2}{2m^2}(\vec E_*^2+\vec B_*^2\,)\Big]^2
-\Big[e(\vec B+\frac{m_*}{m}\,\vec B_*)\pm 
\frac{e^2}{m^2}\,\vec E_*\times\vec
B_*\Big]^2=0\, .
\label{crit}
\ee
Observe that for $Y=0=Z$, {\it i.e.}, $m_*=m$, and with $\F^{\m\n}$ 
growing unboundedly for large $\F^{\m\n}$, the model is not causal:
For $\F^{\m\n}=0$ the expression~\eqn{crit} is positive
but the first term in square brackets must have a zero for large enough 
$\vec E_*$ or $\vec B_*$ at which point the second term is necessarily
negative, and solutions to~\eqn{crit} will exist.
While this shows that broad classes of generalised couplings remain acausal,
the above completely general criterion can applied to a systematic search for
causal models.

\end{appendix}


\begin{thebibliography}{99} 

\bibitem{Weinberg:1970bu}
S.~Weinberg in ``Lectures on Elementary Particles and Quantum Field
Theory'', Volume 1, Brandeis University Summer Institute 1970
(S.~Deser, M.~Grisaru and H.~Pendleton, 
editors, M.I.T. Press, Cambridge, 1970)

\bibitem{Belinfante:1953}
F.~Belinfante,
``Intrinsic Magnetic Moment of Elementary Particles of Spin 3/2,''
Phys.\ Rev.\ {\bf 92}, 997 (1953);
K.~Case,
``The Nonrelativistic Limit of Half-Integral Spin-Wave Equations,''
Phys.\ Rev.\ {\bf 94}, 1442 (Z6) (1954).

\bibitem{Cornwall:1974km}
J.~M.~Cornwall, D.~N.~Levin and G.~Tiktopoulos,
``Derivation of Gauge Invariance from High-Energy Unitarity Bounds on the S - Matrix,''
Phys.\ Rev.\  {\bf D10}, 1145 (1974);
C.~H.~Llewellyn Smith,
``High-Energy Behavior and Gauge Symmetry,''
Phys.\ Lett.\  {\bf B46}, 233 (1973).

\bibitem{Ferrara:1992yc}
S.~Ferrara, M.~Porrati and V.~L.~Telegdi,
``g = 2 as the Natural Value of the Tree Level Gyromagnetic Ratio of 
Elementary Particles,''
Phys.\ Rev.\  {\bf D46}, 3529 (1992).

\bibitem{Johnson:1961vt}
K.~Johnson and E.~C.~Sudarshan,
``Inconsistency of the Local Field Theory of Charged Spin 3/2 Particles,''
Annals Phys.\  {\bf 13}, 126 (1961)

\bibitem{Velo:1969bt}
G.~Velo and D.~Zwanziger,
``Propagation and Quantization of Rarita-Schwinger 
Waves in an External Electromagnetic Potential,''
Phys.\ Rev.\  {\bf 186}, 1337 (1969).

\bibitem{Freedman:1977aw}
D.~Z.~Freedman and A.~Das,
``Gauge Internal Symmetry in Extended Supergravity,''
Nucl.\ Phys.\  {\bf B120}, 221 (1977).

\bibitem{Deser:1977uq}
S.~Deser and B.~Zumino,
``Broken Supersymmetry and Supergravity,''
Phys.\ Rev.\ Lett.\  {\bf 38}, 1433 (1977).

\bibitem{Deser:2000}
S.~Deser, V.~Pascalutsa, M.~Porrati and A.~Waldron, in progress.

\bibitem{Ferrara:1976fu}
S.~Ferrara and P.~van Nieuwenhuizen,
``Consistent Supergravity with Complex Spin 3/2 Gauge Fields,''
Phys.\ Rev.\ Lett.\  {\bf 37}, 1669 (1976).

\bibitem{Low:1954kd}
F.~E.~Low,
``Scattering of Light of Very Low Frequency by Systems of Spin 1/2,''
Phys.\ Rev.\  {\bf 96}, 1428 (1954);
M.~Gell-Mann and M.~L.~Goldberger,
``Scattering of Low-Energy Photons by Particles of Spin 1/2,''
Phys.\ Rev.\  {\bf 96}, 1433 (1954);

\bibitem{Lapidus:1961}
L.~Lapidus and C.~Kuang-Chao,
``The Elastic Scattering of $\gamma$ Rays by Deuterons Below the
Pion-Production Threshold,''
JETP, {\bf 12}, 898 (1961);
K.~Bardakci and H.~Pagels,
``Low Energy Theorems for Photon Processes,''
Phys.\ Rev.\ {\bf 166}, 1783 (1968);
S.~Saito,
``Low-Energy Theorem for Compton Scattering,''
Phys.\ Rev.\ {\bf 184}, 1894 (1969).

\bibitem{Auvil:1966}
P.~Auvil and J.~Brehm, 
``Wave Functions for Particles of Higher Spin''
Phys.\ Rev.\ {\bf 145}, 1156 (1966); 
T.~Moroi,
``Effects of the Gravitino on the Inflationary Universe,''
hep-ph/9503210.

\bibitem{Vermaseren:1991}
Elementary diagrams are rapidly computed with the algebraic
manipulation program Form:
J.~A.~M. Vermaseren,
``Symbolic Manipulation with FORM, version 2''
Computer Algebra Nederland, Amsterdam, 1991.

\bibitem{Piccinini:1984dd}
F.~Piccinini, G.~Venturi and R.~Zucchini,
``Acausality and Violation of S Matrix Unitarity for 
Rarita-Schwinger Particles in an External Electromagnetic Potential,''
Lett.\ Nuovo Cim.\  {\bf 41}, 536 (1984).

\bibitem{Hagen:1982ez}
C.~R.~Hagen and L.~P.~Singh,
``Search for Consistent Interactions of the Rarita-Schwinger Field,''
Phys.\ Rev.\  {\bf D26}, 393 (1982).

\bibitem{Dirac}
Details of the canonical formulation for spin 3/2 may be found in:
A.~Hasumi, R.~Endo and T.~Kimura,
``Dirac Quantization of 
Massive Spin 3/2 Particle Coupled with Magnetic Field,''
J.\ Phys.\ A {\bf A12}, L217 (1979);
K.~Inoue, M.~Omote and M.~Kobayashi,
``Quantization of a Spin 3/2 Field Interacting with the 
Electromagnetic Field,''
Prog.\ Theor.\ Phys.\  {\bf 63}, 1413 (1980);
M.~Yamada,
``Path Integral of Constrained Fermionic Oscillators and its
Application 
to the Interacting Spin 3/2 Field,''
Nuovo Cim.\  {\bf 91A}, 205 (1986);
W.~Cox,
``On the Lagrangian and Hamiltonian Constraint Algorithms for the 
Rarita-Schwinger Field Coupled to an External Electromagnetic Field,''
J.\ Phys.\ A {\bf A22}, 1599 (1989);
V.~Pascalutsa,
``Quantization of an Interacting Spin-3/2 Field and the Delta Isobar,''
Phys.\ Rev.\  {\bf D58}, 096002 (1998)
[hep-ph/9802288].

\bibitem{Madore:1973}
J.~Madore and W.~Tait,
``Propagation of Shock Waves in Higher Spin Wave Equations,''
Commun.\ Math.\ Phys.\ {\bf 30}, 201 (1973).

\bibitem{Madore:1975}
J.~Madore,
``The Characteristic Surfaces of a Classical Spin 3/2 Field 
in an Einstein--Maxwell Background''
Phys.\ Lett.\ {\bf B55}, 217 (1975).

\bibitem{Singh:1974qz}
L.~P.~Singh and C.~R.~Hagen,
``Lagrangian Formulation for Arbitrary Spin. 1. The Boson Case, 
2. The Fermion Case,''
Phys.\ Rev.\  {\bf D9}, 898, 910 (1974);

\bibitem{Schwinger:1970}
J.~Schwinger,
``Particles, Sources, and Fields'', Volume I ~(Addison--Wesley, 1970).


\end{thebibliography}
\end{document}